\documentclass[preprint,12pt]{aastex}

\shorttitle{Kinematics of Tidal Debris}
\shortauthors{Mizutani et~al.}

\begin{document}

\title{Kinematics of Tidal Debris from Omega Centauri's
       Progenitor Galaxy}

\author{Arihiro Mizutani\altaffilmark{1},
        Masashi Chiba\altaffilmark{2},
    and Tsuyoshi Sakamoto\altaffilmark{1}}

\altaffiltext{1}{Department of Astronomical Science, The Graduate University
for Advanced Studies, Mitaka, Tokyo 181-8588, Japan;
mizutnar@cc.nao.ac.jp}
\altaffiltext{2}{National Astronomical Observatory, Mitaka, Tokyo 181-8588,
Japan}

\begin{abstract}
We present the kinematic properties of a tidally disrupted dwarf galaxy in the
Milky Way, based on the hypothesis that its central part once contained the
most massive Galactic globular cluster, $\omega$ Cen. Dynamical evolution of
a self-gravitating progenitor galaxy that follows the present-day and likely
past orbits of $\omega$ Cen is calculated numerically and the kinematic nature
of their tidal debris is analyzed, combined with randomly generated stars
comprising spheroidal halo and flat disk components. We show that the retrograde
rotation of the debris stars at $\sim -100$ km~s$^{-1}$ accords with a recently
discovered, large radial velocity stream at $\sim 300$ km~s$^{-1}$ towards the
Galactic longitude of $\sim 270^\circ$. These stars also contribute, only in
part, to a reported retrograde motion of the outer halo at the North Galactic
Pole. The prospects for future debris searches and the implications for the
early evolution of the Galaxy are briefly presented.
\end{abstract}


\section{INTRODUCTION}

Omega Centauri, the most massive globular cluster in the Milky Way,
is unique in terms of its metallicity content, internal kinematics, and
structure. Unlike other Galactic globular clusters, $\omega$ Cen shows
a wide spread in metallicity (e.g. Norris, Freeman,
\& Mighell 1996), with a main metal-poor component at [Fe/H]$\simeq -1.6$,
a second smaller peak at [Fe/H]$\simeq -1.2$, and a long tail extending
up to [Fe/H]$\simeq -0.5$. The metal-rich population holds
a low velocity dispersion and no sign of rotation, in contrast to the rotating
metal-poor population.
Furthermore, the metal-rich stars in $\omega$ Cen are largely enhanced
in $s$-process elements relative to those in globular clusters and
field stars with similar metallicities (e.g. Norris \& Da~Costa 1995),
thereby suggesting that the ejecta from low-mass,
asymptotic giant branch (AGB) stars had to be retained and incorporated
into the next-generation stars.

In spite of its large mass ($5\times 10^6$ M$_\odot$), it has been
demonstrated by Gnedin et al. (2002) that $\omega$ Cen is not unique in
its ability to retain the AGB ejecta as found for other
clusters. An isolated formation of $\omega$ Cen is thus unlikely, because the
enriched gas would easily be lost by encountering the Galactic disk. The most
viable explanation for the uniqueness of $\omega$ Cen is that it was once
the dense nucleus of a dwarf galaxy (Freeman 1993). A gravitational
potential provided by progenitor's stellar system and dark matter
(as suggested from dwarf spheroidal galaxies in the Local Group, Mateo 1998)
would help retaining the enriched gas and let the cluster being self-enriched
at least over a few Gigayears.

If this hypothesis is the case for the origin of $\omega$ Cen, the question
arises: {\it where and in what form does the stellar system of its progenitor
galaxy remain?} Dinescu (2002) first investigated this issue,
by examining the possible signature of the progenitor's tidal debris among
nearby metal-poor stars in the catalog of Beers et al. (2000, B00).
She identified a group of stars with $-2.0<$[Fe/H]$\le-1.5$, which
departs from the characteristics of the inner Galactic halo but has
retrograde orbits similar to $\omega$ Cen. Her simplified disruption model of
the progenitor galaxy demonstrated that trailing tidal debris, having
orbital characteristics similar to the cluster, can be found in the solar
neighborhood, although the concrete spatial distribution and kinematics
of the debris stars remain yet unclear.

This work motivates us to conduct an N-body simulation for the tidal
disruption of $\omega$ Cen's progenitor galaxy, to obtain the
characteristic structure and kinematics of its debris stars and compare
with various observations showing signatures of recent merging events
in the Milky Way (Gilmore, Wyse, \& Norris 2002, GWN; Kinman et al. 2002,
K02; Chiba \& Beers 2000, CB).
In particular, we show that a recently identified stream of stars at
heliocentric radial velocity of $\sim 300$ km~s$^{-1}$ (GWN) is a natural
outcome of the current disruption model, without affecting
local halo kinematics near the Sun and microlensing optical depth towards
the Large Magellanic Cloud (LMC).

\section{SIMPLE DYNAMICAL MODEL OF A PROGENITOR GALAXY}

We calculate the dynamical evolution of an orbiting dwarf galaxy in a fixed
external gravitational potential representing the Milky Way. The potential
consists of three parts: a spherical Hernquist bulge
$\Phi_b(r)$, a Miyamoto-Nagai disk $\Phi_d (R,z)$, and a logarithmic
dark halo $\Phi_h(r)$, where $r$ is the Galactocentric distance and
$(R,z)$ denote cylindrical coordinates. Each is given as,
$\Phi_b (r) = - GM_b/(r+a)$, $\Phi_d(R,z) =
- GM_d/\sqrt{R^2+(b+\sqrt{z^2+c^2})^2}$,
and $\Phi_h(r) = v_h^2 /2 \ln(r^2+d^2)$,
where $M_b=3.4\times 10^{10}$ M$_\odot$, $a=0.7$ kpc,
$M_d=10^{11}$ M$_\odot$, $b=6.5$ kpc, $c=0.26$ kpc, $v_h=186$ km~s$^{-1}$,
and $d=12$ kpc. This choice yields a circular velocity
of 228 km~s$^{-1}$ at the solar circle of $R_\odot=8$ kpc and a flat
rotation curve outside $R_\odot$.

We set self-gravitating particles in the dwarf galaxy following a King
model, where the central density, central velocity dispersion, and core radius
are given as 0.3 M$_\odot$~pc$^{-3}$, 18.1 km~s$^{-1}$, and 0.56 kpc,
respectively. In addition, a particle with the mass of $5 \times 10^6$
M$_\odot$ representing $\omega$ Cen is placed at the center of the
galaxy to trace its orbit. This setting yields the total mass of the galaxy as
$M_{tot}=5.79 \times 10^8$ M$_\odot$. A part of the mass is provided by
stars, which is roughly estimated from the mean metallicity of
stars in $\omega$ Cen ($\langle$[Fe/H]$\rangle\sim-1.6$), combined
with the metallicity-luminosity relation for the Local Group dwarfs
(C\^{o}t\'{e} et al. 2000) and the mass-to-light ratio (assuming $M/L \sim 4$
obtained for $\omega$ Cen, Meylan et al. 1995),
yielding $M_{stars}\sim 10^7$ M$_\odot$. Thus, our model galaxy
is largely dominated by a dark component, in agreement with the observed
large $M/L$ in dwarfs (Mateo 1998). The galaxy is
represented by a collection of $10^4$ particles and the self-gravity is
calculated in terms of a multiple expansion of the internal potential to
fourth order (Zaritsky \& White 1988).

In the course of its orbital motion, a dwarf galaxy is disrupted by
Galactic tides, whereas its dense core is expected to survive and follow
$\omega$ Cen's orbit. While our calculation in a fixed Galactic
potential neglects dynamical friction against progenitor's orbit, the effect
is only modest for the system of $\la 10^8$~M$_\odot$ (Zhao 2002), especially
during a few orbital periods required for tidal disruption.
We thus examine two representative orbits for the progenitor, model 1
and 2: model 1 follows the current orbit of $\omega$ Cen, whereas for model 2,
we calculate an orbit back to the past over $\sim 2$ Gyr from its current
position and velocity by fully taking into account dynamical friction
and then set a progenitor galaxy on its non-decaying orbit. These two models
provide us with satisfactory information on the generic properties of a tidally
disrupted progenitor and we postulate that the realistic nature of their debris
is midway between these model predictions.
We calculate $\omega$ Cen's orbit, based on the distance\footnote{We estimate
this distance, based on the Harris (1996) data and the relation between
absolute magnitude of cluster horizontal branch and metallicity derived by
Carretta, Gratton, \& Clemintini (2000).} from the Sun $D=5.3 \pm 0.5$ kpc,
position $(l,b)=(309^\circ,15^\circ)$, proper motion
$(\mu_\alpha \cos\delta,\mu_\delta) =(-5.08\pm0.35, -3.57\pm0.34)$
mas~yr$^{-1}$, and heliocentric radial velocity $v_{los} = 232.5 \pm 0.7$
km~s$^{-1}$ (Dinescu, Girard, \& van Altena 1999).
This orbit for model 1 is characterized by frequent disk crossings with a period
of $\tau_{orb} = 0.8 \times 10^8$ yr, retrograde motion, and apo and pericentric
distances $(r_{apo},r_{peri}) = (6.4, 1.1)$ kpc. For model 2,
we obtain $\tau_{orb} = 1.5 \times 10^8$ yr and
$(r_{apo},r_{peri}) = (11.3, 3.0)$ kpc. 
In both experiments, we adopt the same progenitor mass $M_{tot}$ and place it
at apocenter to maximize its survival chances\footnote{For model 1, this initial
mass may be too large as an as-yet-undisrupted galaxy (Zhao 2002), so the debris
density in model 1 can be overly represented.}.

Figure 1 shows the spatial distribution of the tidally disrupted debris.
Upper (middle) panel shows model 1 (model 2) after the 1.37 (1.86) Gyr orbital
excursion of the progenitor galaxy. Lower panel shows the orbit of the galaxy
center. In the course of the orbital motion of the
galaxy, its structure is made elongated along the orbit induced by Galactic
tides, in particular at its pericenter passages, and then the particles are
spread out to form the tidal streams along the orbit.
A rosette-like feature of the debris becomes steady after about eight orbital
periods. Model 1 results in more compact distribution than model 2, which
reflects the difference in orbital radii.

Figure 2 shows the velocity distributions of the debris particles in
cylindrical coordinates $(v_R,v_\phi,v_z)$.
As is evident, model 1 and 2 provide essentially the same debris kinematics:
most remarkable is a sharply peaked $v_\phi$ distribution at
$\sim -100$ km~s$^{-1}$, arising from a retrograde orbit of a progenitor.
These kinematics suggest that the difference in model 1 and 2 resides only
in the spatial extent of the debris.

\section{EFFECTS OF DEBRIS ON GALACTIC KINEMATICS}

\subsection{Method for kinematic analysis}

In order to assess the reality of the debris stars in light of observed
stellar kinematics in the Milky Way, we analyze the kinematics of
both the simulated debris and other Galactic stars
generated randomly by a Monte Carlo method. The metal-poor halo
is modeled as a flattened spheroid $\rho \propto
(R^2 + z^2/q^2)^{-3.5/2}$, where $q$ is an axis ratio ranging 0.55-0.7,
anisotropic velocity ellipsoid $(\sigma_R,\sigma_\phi,\sigma_z)=(154,121,96)$
km~s$^{-1}$, and small mean rotation $\langle v_\phi \rangle= 24$ km~s$^{-1}$,
as found for halo stars with [Fe/H]$<-2$ near the Sun (CB).
Thin and thick disks are modeled as $\rho \propto
\exp(-R/R_d) \sec^2 (z/z_d)$, where $R_d=3.5$ kpc and $z_d=0.3$ (1)
kpc for thin (thick) disk. Both disks
rotate at 200 km~s$^{-1}$, having velocity ellipsoids of
$(34,25,20)$ km~s$^{-1}$ and $(46,50,35)$ km~s$^{-1}$ for thin and thick
disks, respectively (CB). The relative fraction of each
component is fixed using observed local densities near the Sun,
in such a manner that the halo and thick-disk densities at
$D<1$ kpc are 0.2~\% and 2~\% of the thin-disk density,
respectively (e.g. Yamagata \& Yoshii 1992).

In our model of $\omega$ Cen's progenitor galaxy, the self-gravitating
particles represent both stars and dark matter. We note that a correct estimate
for the fraction of stars is uncertain, because their $M/L$ ratio as well as
the amount of dark matter in the progenitor is unavailable. As a useful method
to incorporate this ambiguity for the current kinematic analysis, we set a
parameter $f$ as the fraction of the debris particles relative to halo stars
near the Sun, {\it when all of the particles are regarded as stellar ones}.
By this, the normalization of the halo density is obtained for the given debris
particles.

A typical value of $f$ for the conversion of the simulated particles to
the stars is estimated in the following manner. Model 1 (model 2) yields
21 (74) particles at $D < 2$ kpc, giving the mass density of
$\rho_g = 0.4 (1.3) \times 10^{-4}$ M$_\odot$~pc$^{-3}$ near
the Sun, whereas the total mass density and
metal-poor halo density have been derived as
$8 \times 10^{-3}$ M$_\odot$~pc$^{-3}$ (Gates, Gyuk, \& Turner 1995) and
$6.4 \times 10^{-5}$ M$_\odot$~pc$^{-3}$ (Gould, Flynn, \& Bahcall 1998),
respectively.
Then, if the debris stars (with $M_{stars} \sim 10^7$ M$_\odot$)
are distributed in the same manner as the simulated particles (with
$M_{tot} =5.74 \times 10^8$ M$_\odot$), which would be a reasonable
approximation in view of the dissipationless nature of stars,
the mass density of the debris stars in the solar neighborhood can be
estimated as $(M_{stars}/M_{tot}) \rho_g = O(10^{-6})$
M$_\odot$~pc$^{-3}$, which is about 1~\% of the halo density.
Thus, $f$, defined here as debris fraction at $D<2$ kpc,
is expected to be of order of a few percents.

\subsection{GWN's radial velocity survey}

Recently, GWN reported a spectroscopic survey of
$\sim 2000$ F/G stars down to $V = 19.5$ mag, in the direction
against Galactic rotation $(l,b)=(270^\circ,-45^\circ)$ and $(270^\circ,
+33^\circ)$, where radial velocities, $v_{los}$, in combination with distances
largely reflect orbital angular momentum. The $v_{los}$ distribution of the
stars a few kpc from the Sun shows two stellar streams
at $v_{los}\sim 100$ km~s$^{-1}$ and $\sim 300$ km~s$^{-1}$, which are not
explained by known Galactic components.
While the stream at $v_{los} \sim 100$ km~s$^{-1}$ was reproduced by
their model of a merging satellite in prograde rotation,
the stream at $v_{los}\sim 300$ km~s$^{-1}$ remains yet unexplained.

Figure 3a shows the $v_{los}$ distribution for the
debris stars of model 2 and halo stars of $q=0.7$ (i.e. without disks)
at $1<D<5$ kpc, $260^\circ<l<280^\circ$, and two fields for $b$.
Figure 3b shows when disk stars are incorporated. As is evident,
the debris stars from $\omega$ Cen's progenitor form a local peak
at $v_{los}\sim 300$ km~s$^{-1}$, which is provided by many stars having
$v_\phi \sim -100$ km~s$^{-1}$. This is in good
agreement with the $v_{los}\sim 300$ km~s$^{-1}$ stream discovered by GWN.
A more flattened halo than the case $q=0.7$ yields a higher peak,
since the density contrast of
the debris relative to halo is made higher in this survey region.
It is worth noting that model 1 yields essentially the same $v_{los}$
distribution as model 2, reflecting the same velocity distribution,
although to attain the same peak height at $\sim 300$ km~s$^{-1}$,
$f$ be a few factor larger and the selected range of $l$ be a few degree
higher than the respective values in model 2, because of less number of
debris stars near the Sun. This rule applies to other considerations
below as well.

\subsection{Kinematics at the North Galactic Pole}

Majewski (1992) suggested that the outer halo at the North Galactic Pole (NGP)
shows a retrograde rotation $\langle v_\phi \rangle \simeq -55$ km~s$^{-1}$
at $z>4$ kpc. Also, K02 reported that their sample of
horizontal branch stars at $2<z<12$ kpc shows a retrograde
rotation at $\langle v_\phi \rangle \simeq -65$ km~s$^{-1}$.
On the other hand, halo stars near the Sun 
show no retrograde rotation (CB).

To investigate the role of the debris stars in this issue, we select
those of model 2 and randomly generated stars at $b>70^\circ$ and
$2<D<5$ kpc (resembling K02's selection). Since the observational determination
of full space velocities involves rather
inaccurate information of proper motions compared to radial velocities,
we convolve the velocity distribution of stars with a Gaussian
distribution for velocity errors, having 1~$\sigma$ of a typical
30 km~s$^{-1}$ error.
The resulting velocity distribution shows a non-Gaussian feature owing to
the presence of the debris stars: the $v_\phi$ distribution holds an extra
peak at $\sim -100$ km~s$^{-1}$ in addition to the
$v_\phi \sim 20$ km~s$^{-1}$ peak, where the former amplitude becomes
comparable to the latter one at $f$ of a few percents,
whereas for $v_R$ and $v_z$, the velocity distributions are made slightly
asymmetric. However, the change of $\langle v_\phi \rangle$ by the inclusion
of the debris stars with $f=5$~\% amounts to only $-19$ ($-14$) km~s$^{-1}$
for $q=0.55$ ($0.7$), which are still insufficient for explaining
the reported $\langle v_\phi \rangle = -35 \sim -65$ km~s$^{-1}$.
Also, if we extend the selection of the stars at higher $z$ or instead consider
model 1, the changes of $\langle v_\phi \rangle$ become smaller than the above
mentioned values, because there are no debris stars in our current model.
Thus, it is safe to conclude that the debris stars
contribute only in part to a reported retrograde motion at the NGP.

\subsection{Local halo kinematics and microlensing towards LMC}

We select the nearby debris and halo stars at $6.5<R<9.5$ kpc, $z<4$~kpc,
and $D<4$ kpc (as was drawn by CB), convolve the velocities
with a Gaussian error distribution of $1~\sigma = 30$ km~s$^{-1}$, and
compare with the corresponding stars with [Fe/H]$\le-2$ in B00.
It follows that the non-Gaussian feature in velocities
is much weaker than that at the
NGP: the change of $\langle v_\phi \rangle$ for $f=5$~\% is only
$-9$ ($-10$) km~s$^{-1}$ in model 1 (model 2). This is due to
the smaller debris fraction near $z=0$ than at high $|z|$.

The effects of the debris stars on the microlensing optical depth towards LMC,
$\tau$, are modest as well. Following the Gould (1999) prescription for $\tau$
and investigating the debris within $10^\circ \times 10^\circ$
centered at LMC, we arrive at $\tau \la 10^{-7}f$, thereby indicating
that $\tau$ provided by the debris stars is much smaller than the observed
$\tau$ of $O(10^{-7})$.

\section{DISCUSSION}

We have demonstrated that our fiducial models of an orbiting dwarf galaxy
that once contained $\omega$ Cen predict a sequence of tidal streams
in retrograde rotation and their existence is imprinted in kinematics of
nearby stars, especially in the direction against Galactic rotation (GWN) and
at the NGP (K02), while local halo kinematics and microlensing towards LMC
remain unchanged. The streams are mostly distributed
inside the solar circle, as suggested from the current orbit of
$\omega$ Cen (Dinescu 2002).
In contrast to the Sgr dwarf galaxy having polar orbit,
the orbit of $\omega$ Cen's progenitor galaxy is
largely affected by a non-spherical disk potential, where the orbital plane
exhibits precession with respect to the Galactic Pole, causing self-crossing
of tidal streams in the disk. The projection of the orbit
perpendicular to the disk shows an 'X'-like feature, thereby leaving
denser streams at high $|z|$ than at low $|z|$ for a given radius.
This explains the significance of the debris at the NGP
compared to the solar neighborhood.

Existing kinematic studies of Galactic stars to search for a signature
of $\omega$ Cen's progenitor galaxy are yet confined to nearby stars,
where the significance of the debris streams is modest, as shown here.
Searches of stars inside the solar circle are more encouraging (Fig. 1),
in particular in the directions of $l \sim 320^\circ$ and $l \sim 50^\circ$,
where we expect the presence of high-velocity streams at
$v_{los} = 200 \sim 300$ km~s$^{-1}$ and $-400 \sim -300$ km~s$^{-1}$,
respectively. Future radial velocity surveys of these fields including the
sample of the Sloan Digital Sky Survey or planned RAdial Velocity
Experiment are worth exploring in this context. Also, detailed
abundance studies of candidate stream stars will be intriguing, because such
stars may exhibit different abundance patterns from field halo stars,
as found in dwarf galaxies (Shetrone, C\^{o}t\'{e}, \& Sargent 2001).

A yet unsettled issue is the origin of a progenitor satellite
orbiting inside the solar circle, because dynamical friction alone from the
present-day {\it smooth} Galactic components is insufficient for shrinking the
orbit if it was born at a large distance (say, $\sim 50$ kpc) from Galactic
center (Zhao 2002). One of the possibilities to preclude it
may be that the merging of a satellite occurred while
the Milky Way was still in the process of halo formation via
hierarchical merging of several subgalactic clumps; successive gravitational
interaction among clumps may help reducing the orbital angular momentum of
a progenitor efficiently. Also, a progenitor may have formed
in the vicinity of the proto-Galaxy, where the environment of a strong tidal
field promotes the formation of a compact M32-like galaxy (Burkert 1994)
and its high density affords the survival chances until
the epoch of the Galactic disk formation. Then, if a progenitor retained
gas, growing Galactic tides induce the infall of gas into the
progenitor center and trigger the formation of a globular cluster
there (Bekki \& Chiba 2002) similar to $\omega$ Cen. The story is yet
speculative but worth pursuing based on sophisticated numerical codes for
Galaxy formation.

\acknowledgments
We are grateful to the anonymous referee for helpful comments.
M.C. thanks Kenji Bekki and Tim Beers for useful discussions.


\clearpage


\clearpage
\begin{figure}
\caption{
Upper (middle) panel shows the spatial distribution of the tidally disrupted
debris for model 1 (model 2) after the 1.37 (1.86) Gyr orbital excursion of
$\omega$ Cen's progenitor galaxy. Lower panel shows the orbit of the galaxy
center for model 1 (dotted line) and 2 (solid line). The plots are projected
onto three orthogonal planes, where the Sun is located at $x=-8$ kpc and $xy$
corresponds to the disk plane. A frame measures 15 kpc on a side of each panel.
The current position of $\omega$ Cen is at $(x,y,z)= (-4.8,-4.0,1.4)$ kpc.
}
\end{figure}

\begin{figure}
\caption{
Velocity distribution of the debris particles in cylindrical coordinates,
$v_R$ (dotted), $v_\phi$ (solid), $v_z$ (dashed), for
model 1 (a) and 2 (b).
}
\end{figure}

\begin{figure}
\caption{
(a) Distribution of the heliocentric radial velocities in the direction of
the GWN survey, for the debris stars of model 2
and randomly generated halo stars with $q=0.7$.
We select the stars at $1<D<5$ kpc and $260^\circ<l<280^\circ$
in the fields of $30^\circ<|b|<50^\circ$ (with $f=1$~\% and 3~\% for
dotted and solid histograms, respectively) and $20^\circ<|b|<40^\circ$
(with $f=5$~\%: dashed histogram).
(b) The same as (a) but incorporating the randomly generated disk
stars as well for $30^\circ<|b|<50^\circ$ with $f=5$~\% (solid histogram).
Dotted histogram denotes the contribution from the metal-poor halo alone.
}
\end{figure}

\end{document}